\documentclass{article}

\usepackage{braket}
\usepackage[english]{babel}

\usepackage[letterpaper,top=2cm,bottom=2cm,left=3cm,right=3cm,marginparwidth=1.75cm]{geometry}

\usepackage{amsmath}
\usepackage{graphicx}
\usepackage{amssymb}
\usepackage{braket}
\usepackage{float}
\usepackage{svg}
\usepackage{nicefrac}
\usepackage{authblk}
\usepackage{hyperref}
\usepackage{mathtools}
\usepackage{comment}
\usepackage[symbol]{footmisc}
\DefineFNsymbols{sym}{ {\ensuremath{\dagger}}  {\ensuremath{*}} {\ensuremath{\diamond}}}
\setfnsymbol{sym}

\title{ 
An entangled photon source for the telecom C-band based on a semiconductor-confined spin\\
}
\author[1,2,$\dagger$,$*$]{Petros Laccotripes}
\author[1,$\dagger$]{Junyang Huang}
\author[1,$\dagger$]{Ginny Shooter}
\author[1]{Andrea Barbiero}
\author[1]{Matthew S. Winnel}
\author[2]{David A. Ritchie}
\author[1]{Andrew J. Shields}
\author[1,$\diamond$]{Tina M\"{u}ller}
\author[1]{R. Mark Stevenson}
\affil[1]{Toshiba Europe Limited, 208 Cambridge Science Park, Milton Road, Cambridge, CB4 0GZ, UK}
\affil[2]{Cavendish Laboratory, University of Cambridge, JJ Thomson Avenue, Cambridge, CB3 0HE, UK}
\date{}

\usepackage{setspace}
\begin{document}
\maketitle
\begin{abstract}
\noindent Multiphoton entangled states are a key resource for quantum networks and measurement-based quantum computation. Scalable protocols for generating such states using solid-state spin-photon interfaces have recently emerged, but practical implementations have so far relied on emitters operating at short wavelengths, incompatible with low-loss fibre transmission. Here, we take a key step towards the generation of telecom wavelength multi-qubit entangled states using an InAs/InP quantum dot. After establishing that all essential criteria for generating cluster states using a ground state spin as the entangler are satisfied, we implement a scalable protocol to entangle the resident spin with sequentially emitted photons directly in the telecom C-band. We demonstrate a two-qubit (spin-photon) entanglement fidelity of $59.5\pm 8.7\%$ and a lower bound of three-qubit (spin-photon-photon) entanglement fidelity of $52.7\pm 11.4\%$. Our results close the performance gap between short-wavelength quantum dot systems and the existing telecom infrastructure, establishing a route towards practical large photonic cluster states for fibre-based quantum network applications.
\end{abstract}

\footnotetext[1]{These authors contributed equally.}
\footnotetext[2]{\textit{Email:} \href{mailto:petros.laccotripes@toshiba.eu}{petros.laccotripes@toshiba.eu}, \href{mailto:pl490@cam.ac.uk}{pl490@cam.ac.uk}}
\footnotetext[3]{\textit{Email:} \href{mailto:tina.muller@toshiba.eu}{tina.muller@toshiba.eu}}



\newpage

\section*{Introduction}

\noindent Entanglement is a defining feature of quantum mechanics and a foundational resource for quantum technologies, enabling capabilities such as quantum computation, secure communication, and quantum networking \cite{wehner2018quantum}. Among the various forms of entanglement, photonic cluster states \cite{Hein2004Multiparty,aghaee2025scaling}, comprising multiple entangled photons, play a central role in many quantum information protocols. Their built-in redundancy enables fault-tolerant operations and enhanced resilience to photon loss, making them especially valuable for measurement-based quantum computation \cite{Varnava2006Loss} and all-optical, memoryless quantum repeater schemes \cite{Azuma2015All, buterakos2017deterministic}. 

Despite their conceptual appeal, the development of efficient photonic cluster state generation remains highly challenging. Sources based on spontaneous parametric downconversion have been demonstrated \cite{walther2005experimental,zhang2006experimental,zhong201812}, but cluster state size is intrinsically limited by a trade-off between brightness and suppression of multi-photon emission. Also, the fusion of small cluster states to create larger states is inherently inefficient \cite{thomas2024fusion, zhong201812}. Although protocols for scalable cluster state generation were proposed theoretically several years ago \cite{lindner2009proposal, lee2019quantum}, experimental realisations have only recently been achieved, using either quantum dot (QD) \cite{schwartz2016deterministic, appel2022entangling, cogan2023deterministic, coste2023high, su2024continuous, meng2024deterministic, huet2025deterministic} or atomic platforms \cite{yang2022sequential, thomas2022efficient}. These approaches typically rely on coherent control of a resident spin, which interacts with a driving laser field and is periodically manipulated to sequentially emit entangled photons. However, demonstrations to date have employed emitters operating at short wavelengths which are incompatible with the low-loss telecom window of optical fibres. Addressing this mismatch is vital for practical quantum communication protocols. While frequency conversion approaches have been successfully employed for few-photon technologies \cite{zaske2012visible, albrecht2014waveguide}, current external conversion efficiencies around 50\% quickly render use of shorter-wavelength systems unfeasible. Native emission of telecom wavelength multi-photon entangled states is crucial for fibre-based quantum network applications, such as all-photonic quantum repeaters \cite{Azuma2015All}.

Here, we demonstrate the generation of multi-qubit entanglement in the telecom C-band using an InAs/InP QD. This platform, along with telecom-optimised InAs/GaAs quantum dot systems \cite{sittig2022thin}, has evolved significantly, progressing from the first reports of single and entangled photon emission, to advanced capabilities such as optical spin control \cite{dusanowski2022optical} and quantum state tomography \cite{laccotripes2024spin, peniakov2025initialization}. Building on recent demonstrations of spin-photon entanglement at telecom wavelengths \cite{laccotripes2024spin, Uysal2025Spin}, we implement a protocol that generates two- and three-qubit entangled states via sequential photon emission from a periodically driven resident spin. This work constitutes the first realisation of a scalable cluster-state generation protocol using quantum emitters compatible with the telecom infrastructure, and represents a major step toward closing the performance gap with shorter-wavelength systems.

This article is organised as follows: we first introduce our QD system and the protocol we use for multiqubit entanglement. We then investigate the coherence times and Land\'e g-factors of two candidate spin states, that of the electron and hole, to select a suitable entangler. We continue on to perform three-photon correlation measurements to demonstrate entanglement between the hole spin and telecom wavelength photons. Finally, we provide a discussion of the limitations of the current system by comparing the results to numerical simulations of the entanglement process, and we give an outlook of the performance expected from an improved QD device.

\section*{Results}
\subsection*{Telecom C-band entangled photon source} 

\begin{figure}[H]
\centering
\includegraphics[width=1\textwidth]{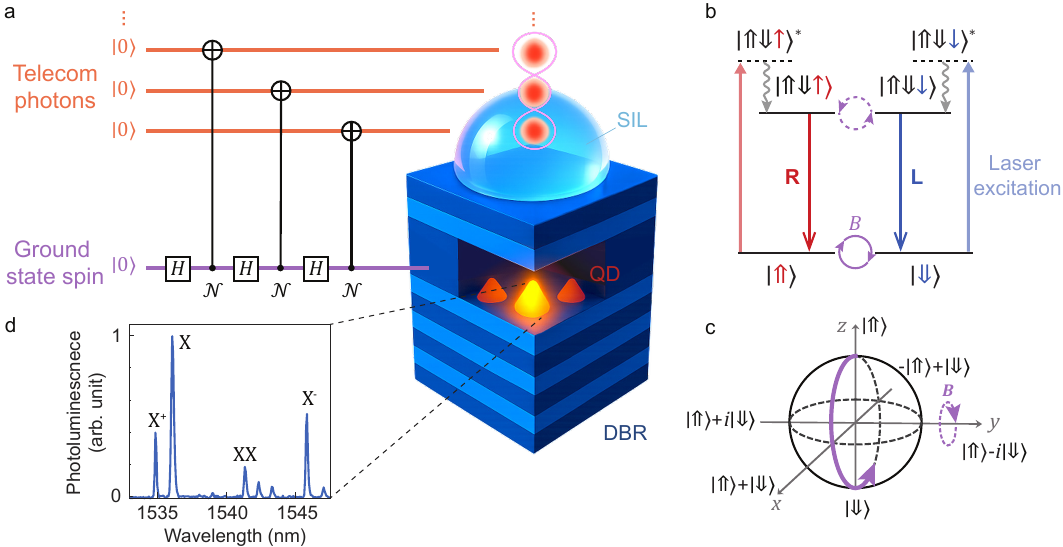}
\caption{\textbf{Telecom C-band entangled photon source.} (a) Circuit diagram of Lindner-Rudolph protocol for one dimensional cluster state generation. The Hadmard gate, $H$, implements spin rotation by 90${}^\circ$, and the CNOT gate, $\mathcal{N}$, realises the spin preserving emission and entanglement generation. (b) Energy level diagram of a positively charged trion system under longitudinal acoustic phonon-assisted excitation showing optical selection rules. (c) Ground state spin precession under the influence of an in-plane magnetic field depicted on the Bloch sphere. (d) Photoluminescence spectrum at telecom wavelength from the QD emitter under above-band laser excitation. Four distinct emission lines are observed, corresponding to the exciton (X), biexciton (XX), negatively charged exciton (X$^-$), and positively charged exciton (X$^+$).}
\label{fig:intro} 
\end{figure}

Our implementation follows the protocol proposed by Lindner and Rudolph \cite{lindner2009proposal}, where a one-dimensional photonic cluster state is generated by alternating between applying a controlled-NOT (CNOT) gate and a Hadamard gate to the QD spin. In our experiment, this sequence is implemented using a string of precisely timed coherent excitation pulses. Each excitation pulse corresponds to the CNOT operation that entangles the spin with an emitted photon, while spin rotation driven by Larmor precession implements the Hadamard gate. The timing of the excitation pulses relative to the spin precession allows for controlled manipulation of the spin state, enabling the construction of a cluster state with polarisation entanglement, as illustrated in \autoref{fig:intro}(a).

The physical system used to generate multi-photon entanglement is schematically shown in \autoref{fig:intro}(b). We employ a QD-confined ground-state spin as the entangling qubit, which mediates the sequential generation of entangled photons. In the absence of an external magnetic field, the trion state decays via two optical transitions with orthogonal circular polarisations ($\ket{R}$ and $\ket{L}$) to a well-defined ground-state spin. Crucially, the excitation process must be spin-preserving to provide the necessary cyclicity for the targeted protocol, with the optical selection rules connecting the polarisation of the emitted photon and the spin in the ground state,
\begin{equation} \label{eq:selectionRules}
   \ket{\Uparrow} \xleftrightarrow{\ket{R}} \ket{\Uparrow\Downarrow\uparrow}\hspace{1 cm} \ket{\Downarrow} \xleftrightarrow{\ket{L}} \ket{\Uparrow\Downarrow\downarrow}.
\end{equation}
To achieve this, we use longitudinal acoustic phonon-assisted (LA-PA) excitation, in which a blue-detuned laser ($\Delta$$\lambda$ = 1.5 nm from the trion resonance) excites a higher energy state before rapid relaxation to the trion via spin-preserving phonon emission \cite{Coste2023Probing}. 

To enable coherent spin manipulation between laser excitation pulses, we apply a weak in-plane magnetic field in the Voigt configuration, inducing Larmor precession of the ground-state spin, as depicted in \autoref{fig:intro}(c).  A Hadamard gate on the ground state spin can then be realised by waiting a quarter of the spin precession period between sequential excitation pulses,
\begin{equation}\label{eq:Hadamard}
H(\ket{\Uparrow})\longrightarrow \ket{+}=\frac{\ket{\Uparrow}+\ket{\Downarrow}}{\sqrt{2}} \hspace{1cm} H(\ket{\Downarrow})\longrightarrow \ket{-}=\frac{-\ket{\Uparrow}+\ket{\Downarrow}}{\sqrt{2}}.
\end{equation}

The sequential application of CNOT and Hadamard gates in our physical system, as described in \autoref{eq:selectionRules} and \autoref{eq:Hadamard}, respectively, implement the multi-photon entanglement protocol. At the core of this protocol is the resident spin in the ground state, which can either be an electron for X$^-$ or a heavy hole for X$^+$. For our InAs/InP QD, both of these trion transitions can be seen in \autoref{fig:intro}(d) under above-band excitation. We therefore have flexibility in selecting the spin species that serves as the entangler.

\subsection*{Coherent dynamics of electron and hole spins} \label{Sec:ExcitedStateTomo}

To enable high-fidelity generation of multi-photon entangled states with the protocol, a QD-confined spin needs to satisfy stringent criteria: the spin coherence time must exceed the total time needed to carry out the protocol, and the Land\'e g-factor must be much higher for the ground state than the excited state spin. The difference in g-factors is essential to enable ground state spin rotation for the Hadamard gate while reducing unwanted rotation of the excited state spin, which would degrade the achievable entanglement fidelity. Here, we evaluate the spin coherence times and g-factors for the excited state valence band heavy-hole and the conduction band electron in the two species of trion \cite{Coste2023Probing}.

\begin{figure}[H]
\centering
\includegraphics[width=1\textwidth]{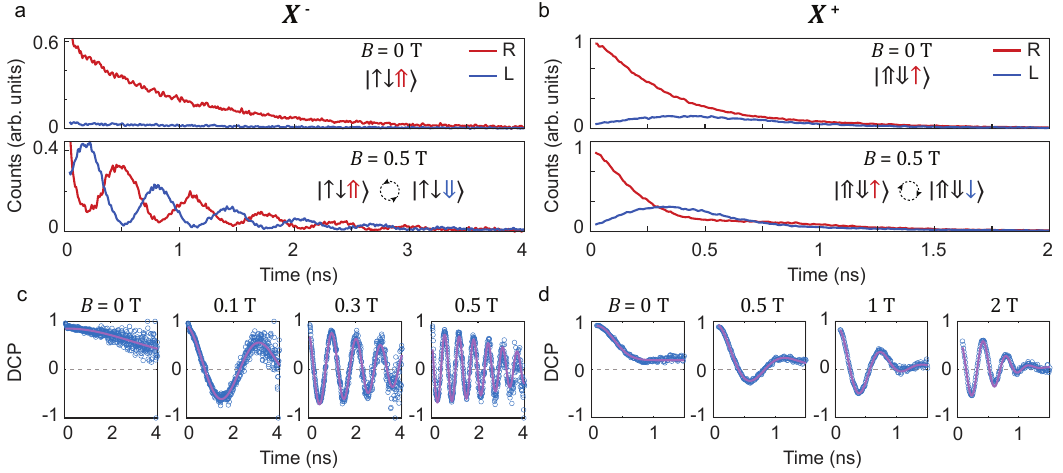} 
\caption{\textbf{Characterisation of an electron and a hole spin in a magnetic field.} Polarisation-sensitive time-resolved photoluminescence emission from the (a) negative and (b) positive trion transitions under 0 T and 0.5 T magnetic field. At 0 T (top panels), the R-polarised emission (dark red) dominates over the L-polarised emission (light blue), as expected by the optical selection rules. (c) and (d) Time-resolved degree of circular polarisation of the PL emission from the X$^-$ and X$^+$, respectively at various magnetic fields. The experimental data (blue circles) is well described by the fits (orange curves); with the oscillation frequency indicating the spin precession, and thus  the g-factor, and the damping envelope quantifying the loss of coherence.}
\label{fig:excited_tomo} 
\end{figure}

We employ a sequence of two laser pulses: an above-band pulse to inject carriers into the QD, followed by an R-polarised LA-PA pulse that spin-selectively populates the trion state. Monitoring the polarisation of the emitted photons allows us to infer the behaviour of the unpaired spin in the excited state during the trion lifetime; a hole for the $\ket{\uparrow\downarrow\Uparrow}$ state of the X$^-$, and an electron for the $\ket{\Uparrow\Downarrow\uparrow}$ state of the X$^+$. \autoref{fig:excited_tomo}(a) and (b) illustrate the time and polarisation resolved measurements of the emission from the X$^-$ and X$^+$ states, respectively. In the absence of a magnetic field (top panels), the time traces show predominantly R-polarised emission, as expected. This leads to a high degree of circular polarisation (DCP), given by $DCP=(I_R-I_L)/(I_R+I_L )$, where $I_R$ and $I_L$ correspond to the intensity of R-polarised and L-polarised light. In contrast to X$^-$, a weak Larmor precession can be observed during the X$^+$ decay. This could be attributed to a residual Overhauser field from the nuclear spin bath. The difference in spin dynamics arises from the nature of carrier-nuclear spin interactions: electrons (s-type orbitals) undergo stronger Fermi contact hyperfine coupling, leading to faster dephasing, while heavy holes (p-type orbitals) interact via weaker dipole-dipole coupling, resulting in slower depolarisation, and therefore are a preferable candidate as the resident entangler spin \cite{testelin2009hole,tartakovskii2007nuclear}.

An external in-plane magnetic field lifts the spin degeneracy and induces a coherent evolution between the spin states via the Zeeman interaction, manifesting in time-dependent oscillations in the R- and L-polarised components in the emission. The Larmor frequency of the precessing spin is given by $f_L=g_s\mu_BB/h$, where $g_s$ is the Land\'e g-factor, $\mu_B$ is the Bohr magneton, $B$ is the magnetic field, and $h$ is Planck’s constant. L-polarised emission can only occur if the spin undergoes a rotation in the excited state within the trion's lifetime, evidenced when the magnetic field is increased to 0.5 T in the bottom panels of \autoref{fig:excited_tomo}(a) and (b).

\autoref{fig:excited_tomo}(c) and (d) present the time-resolved DCP for increasing magnetic fields, which is fitted using
\begin{equation}
    DCP(t) = P_0  e^{-(t/T_2^*)^2} \text{cos}(2\pi f_L t),
\end{equation}
where $P_0$ is the initial degree of circular polarisation. The damping envelope of the oscillation corresponds to the limited coherence time ($T_2^*$) of the spin, and the oscillation frequency depends on the g-factor of the spin. A larger g-factor will lead to a higher oscillation frequency for the same applied magnetic field, which in our case is clearly observed for the hole spin in \autoref{fig:excited_tomo}(c) compared to the electron spin in \autoref{fig:excited_tomo}(d). 

The g-factors of the electron, $g_e=0.096 \pm 0.004$, and the hole, $g_h=0.229 \pm 0.001$, are extracted (see Supplementary Information Section 3), and agree with those reported in literature for InAs/InP QDs \cite{laccotripes2024spin} and are slightly lower than those reported for InAs/GaAs QDs \cite{ramesh2025impact, peniakov2025initialization}. The coherence times of $T_2^* (e)  = 0.8 \pm 0.1$  ns and $T_2^* (h)> 4.8 \pm 0.5$ ns are similar to recently reported values for electron and hole spins in InAs/GaAs telecom wavelength QDs \cite{peniakov2025initialization}. A comparatively longer hole spin coherence time is also commonly observed for QDs emitting at 900 nm \cite{cogan2018depolarization}, and is attributed to its lower hyperfine interaction compared to the electron spin, diminishing the decohering influence of the nuclear spin bath \cite{gerardot2008optical, fischer2008spin, eble2009hole}. It is important to note that in the case of the hole spin coherence measured here, the measurement is limited by the radiative decay time $T_{\text{rad}}^{X^-} \sim$ 1.1 ns and therefore provides only a lower bound. Given the superior spin coherence and higher g-factor of the hole, we proceed to use the positively charged exciton X$^+$ for cluster state generation, with a ground state hole as the entangler.

\subsection*{Multi-qubit entanglement} \label{Sec:MultiQubitResults}

Next, we select the magnetic field at which to conduct our experiment, phenomenologically modelling the scheme as discussed in Supplementary Information Section 4. A low magnetic field will preserve phonon indistinguishability and suppress unwanted precession of the excited state. However, a low field also slows down the implementation of the Hadamard gate, causing the multi-photon entanglement protocol to exceed the coherence time of the ground-state spin. We find a good compromise at a magnetic field of approximately $40$ mT. At this field strength, the hole spin undergoes Larmor precession with a period of $T_h = 2\pi \hbar / (g_h\mu_B B) \approx 8$ ns, which is significantly longer than the radiative decay time of $T_{\text{rad}}^{X^+} \sim$ 0.8 ns. This also corresponds to a high ratio between the Larmor period of the electron and the radiative decay time $T_e/T_{\text{rad}}^{X^+}\sim 24$, ensuring high-fidelity CNOT gate operation.  

\begin{figure}[H]
\centering
\includegraphics[width=0.5\textwidth]{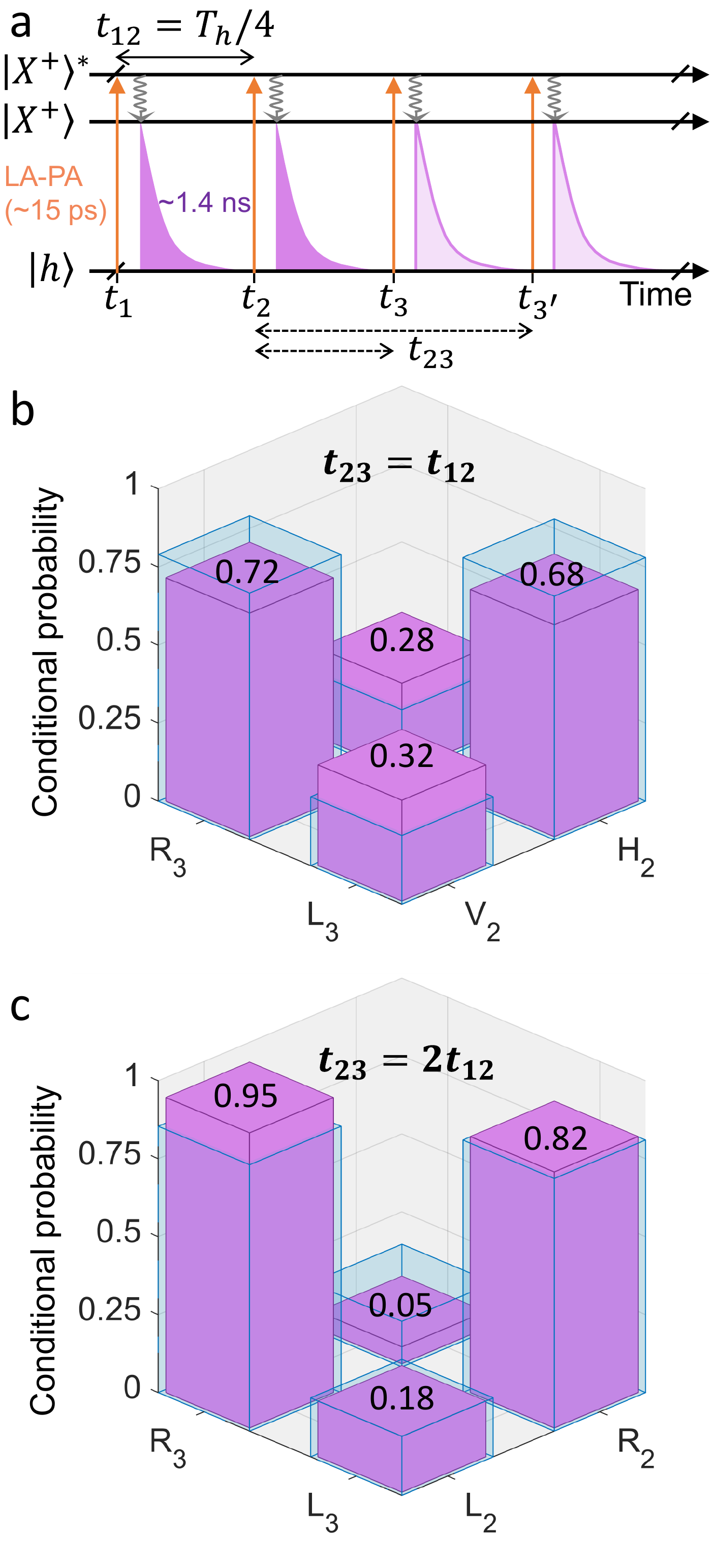} 
\caption{\textbf{Multi-photon entanglement generation.} (a) Excitation pulse sequence using LA-PA pulses (orange arrows) and the emitted single photons (purple decays). Pulse spacing (e.g. $t_{12}$) is set to a quarter of the heavy-hole spin precession period $T_h$. Three consecutive photons are detected, and their polarisations are projected onto the R, L, H, and V bases. The delay $t_{23}$ is either set to $t_{12}$ or ${2t}_{12}$. (b) and (c) Truth tables showing the probability of the polarisation of photon \#2 conditioned on the polarization of photon \#3, for pulse spacing $t_{23}=t_{12}$ and $t_{23}={2t}_{12}$, respectively (solid pink bars). Results from a master equation simulation are overlaid in semi-transparent blue bars.}
\label{fig:entanglement} 
\end{figure}

We then implement the linear cluster state generation protocol, and determine the entanglement fidelity between the spin and emitted photons \cite{coste2023high}. To quantify the degree of entanglement, we perform conditional three-photon correlation measurements, corresponding to spin initialisation, entanglement generation, and spin readout. We use linearly polarised LA-PA pulses, $\ket{H}$ (see Supplementary Information Section 5), to coherently convert any resident hole into the trion manifold with equivalent probability amplitudes for the spin projections. Three pulses excite the QD with respective delays $t_{12}=t_2-t_1$ and $t_{23}=t_3-t_2$, as illustrated in \autoref{fig:entanglement}(a). The pulse spacing $t_{12}$ is set to 2.08 ns ($\sim$480 MHz), for $\mathrm{\pi}$/2 Larmor precession of the ground state hole between the first two pulses. Since the hole coherence time $T_2^* > 4.8$ ns, the spin retains significant coherence throughout the three‐pulse sequence required to produce the linear cluster state. To probe a pair of orthogonal bases for the entanglement measurement, we vary $t_{23}$ to achieve a Larmor precession of $\mathrm{\pi}$/2 and then $\mathrm{\pi}$. 

Following excitation by the first LA-PA pulse and detection of an R-polarised photon, the system is initialised in the $\left|\Uparrow\right\rangle$ state. The spin then precesses around the in-plane magnetic field, evolving into the superposition state $\left|+\right\rangle=\left|\Uparrow\right\rangle+\left|\Downarrow\right\rangle$ as shown in \autoref{fig:intro}(c). A second LA-PA pulse leads to the emission of a second photon, yielding an entangled state between the spin and emitted photon
\begin{equation}
    \ket{\psi_1} = 1/\sqrt{2} (\ket{\Uparrow,R_2} + \ket{\Downarrow,L_2}).
\end{equation}
During the interval $t_{23}= t_{12}$, the spin state evolves from $\ket{+}$ to $\ket{\Downarrow}$. Consequently, after the third excitation pulse and subsequent photon emission, the spin-photon-photon system is ideally projected into a tripartite entangled state
\begin{equation} \label{eq:t12} 
    \ket{\psi_2} = 1/\sqrt{2} (-i\ket{\Uparrow,V_2,R_3} + \ket{\Downarrow,H_2,L_3}).
\end{equation}
Further excitations would lead to the addition of photons to the entangled string, with a phase dependent on the delay between excitation pulses \cite{thomas2022efficient}. To quantify the spin-photon entanglement, it is necessary to measure correlations in an orthogonal bases by setting $t_{23}=2t_{12}$. In this case, the system instead evolves to
\begin{equation} \label{eq:2t12} 
   \ket{\psi_3} = 1/\sqrt{2} (-\ket{\Uparrow,L_2,R_3} + \ket{\Downarrow,R_2,L_3}).
\end{equation}

We now demonstrate spin–photon entanglement using photons \#2 and \#3; analysing their projections in the H/V and R/L bases. The relatively long excited state lifetime leads to unwanted precession of the excited state spin prior to photon emission, which we overcome by implementing a post-selection window of 300 ps starting at the onset of the radiative decay after each excitation. \autoref{fig:entanglement}(b) shows the measured conditional probabilities for state $ \left|\psi_2\right\rangle$ where photon \#2 is measured in the H/V basis and photon \#3 in the R/L basis, projecting the spin state to $\left|\Downarrow\right\rangle$ or $\left|\Uparrow\right\rangle$, respectively. We obtain the following conditional coincidences: $P\left(V_2|L_3\right)=0.32\pm0.15$, $P\left(V_2|R_3\right)=0.72\pm0.23$, $P\left(H_2|L_3\right)=0.68\pm0.18$, and $P\left(H_2|R_3\right)=0.28\pm0.11$. Similarly, \autoref{fig:entanglement}(c) shows the conditional probabilities for state $\left|\psi_3\right\rangle$ where both photons \#2 and \#3 are measured in the circular basis. We obtain: $P\left(L_2|L_3\right)=0.18\pm0.07$, $P\left(L_2{|R}_3\right)=0.95\pm0.21$, $P\left(R_2|L_3\right)=0.82\pm0.19$, and $P\left(R_2|R_3\right)=0.05\pm0.04$. 

In \autoref{fig:entanglement}(b) and (c), we also present simulated results for our system, shown as semi-transparent blue bars, obtained using a master equation approach (see Supplementary Information Section 7). These simulations incorporate previously extracted parameters, including hole and electron spin coherence times, g-factors, and the trion lifetime. The model aligns closely with the experimental data, demonstrating its reliability. In the discussion we extend this simulation to model the generation of larger cluster states. 

We calculate the spin-photon entanglement fidelity, $F_{s,p}=\bra{\psi}\rho\ket{\psi}$, using methods developed in \cite{blinov2004observation,gao2012observation}. We measure the diagonal elements of the density matrix in two bases and use them to calculate the fidelity, $F$, in each basis,

\begin{equation}
\begin{split}
            F_1 &= \rho_{L\Uparrow,L\Uparrow}+ \rho_{R\Downarrow,R\Downarrow} -2\sqrt{\rho_{R\Uparrow,R\Uparrow}* \rho_{L\Downarrow,L\Downarrow}} = 79.0 \pm 7.7 \% \\
            F_2 &= \tilde{\rho}_{V\Uparrow,V\Uparrow} +\tilde{\rho}_{H\Downarrow,H\Downarrow} -\tilde{\rho}_{V\Downarrow,V\Downarrow} -\tilde{\rho}_{H\Uparrow,H\Uparrow} = 40.0 \pm 15.5 \% 
\end{split}
\end{equation}

where $\rho_{ab,ab} =\bra{a,b}\rho \ket{a,b} $ are the diagonal matrix elements in the first measurement basis, and $\tilde{\rho}_{ab,ab}$ in the rotated basis. We obtain the density matrix elements $\rho_{ab,ab} = \frac{1}{2} P(a|b)$ using the conditional probabilities on polarisation $\vec{p}$ and spin $s$ given earlier (see Supplementary Information Section 6). The overall measured spin-photon fidelity is
\begin{equation}
F_{s,p}\geq \frac{F_1 + F_2}{2} = 59.5 \pm 8.7\%.
\end{equation}

Lastly, we use $F_{s,p}$ to derive a fidelity lower bound to the three-qubit spin-photon-photon entangled state under the assumptions that the spin is in a maximally mixed state at the beginning of the experiment and that after initialisation the system would emit two photons of the same polarisation as the time between their emissions tends to zero, $\lim\limits_{t_{12} \to 0} P(\vec{p}_1 = \vec{p}_2)=1$ \cite{coste2023high}. The spin-photon-photon fidelity lower bound is calculated as

\begin{equation}
    F_{s,p,p} =  F_{s,p} \times \eta = 52.7 \pm 11.4\%, 
\end{equation}
where $\eta= 88.5 \pm 14.2 \%$ quantifies the preservation of entanglement with every emission process. $\eta$ was estimated using the three-photon coincidences in the circular basis, which is less sensitive to polarisation misalignment in our detection system. It should be noted that $F_{s,p}$ and $F_{s,p,p}$ are lower bounds as they are obtained through correlation measurements after additional spin decoherence that has occurred between $t_{23} = t_{12}$ and $t_{23} = 2t_{12}$. Therefore, these bounds are not tight, as they tend to overestimate the impact of spin dephasing in their calculation.

\section*{Discussion}

In the derivation of the ideal entangled states given in \autoref{eq:t12} and \autoref{eq:2t12}, we neglected any experimental imperfections such as non-ideal initialisation, decoherence processes, or timing jitters, though in practise each of these effects will reduce the achievable entanglement fidelity. Here, we investigate the limitations of our current experiment and QD device, and consider the future potential with pragmatic improvements. We develop a numerical model of our QD device using a Lindblad master equation (see Supplementary Information Section 7), and implement polarisation-resolved photon-number decomposition to extract more realistic fidelity estimates $F_{s,p}$ and $F_{s,p,p}$. Furthermore, we carried out a full quantum process tomography simulation, using the Pauli transfer matrix formalism. We simulate a single cycle of the Lindner and Rudolph protocol where a spin–photon CNOT gate models the photon emission step and a subsequent Hadamard gate represents the spin rotation. From this we obtain the process map, $\mathcal{E}(\rho)$, which characterises the combined emissions and rotation operations, and enables direct computation of the joint spin--$k$-photon density matrix after $k$ repetitions of this step. 

\begin{figure}[H]
\centering
\includegraphics[width=0.6\textwidth]{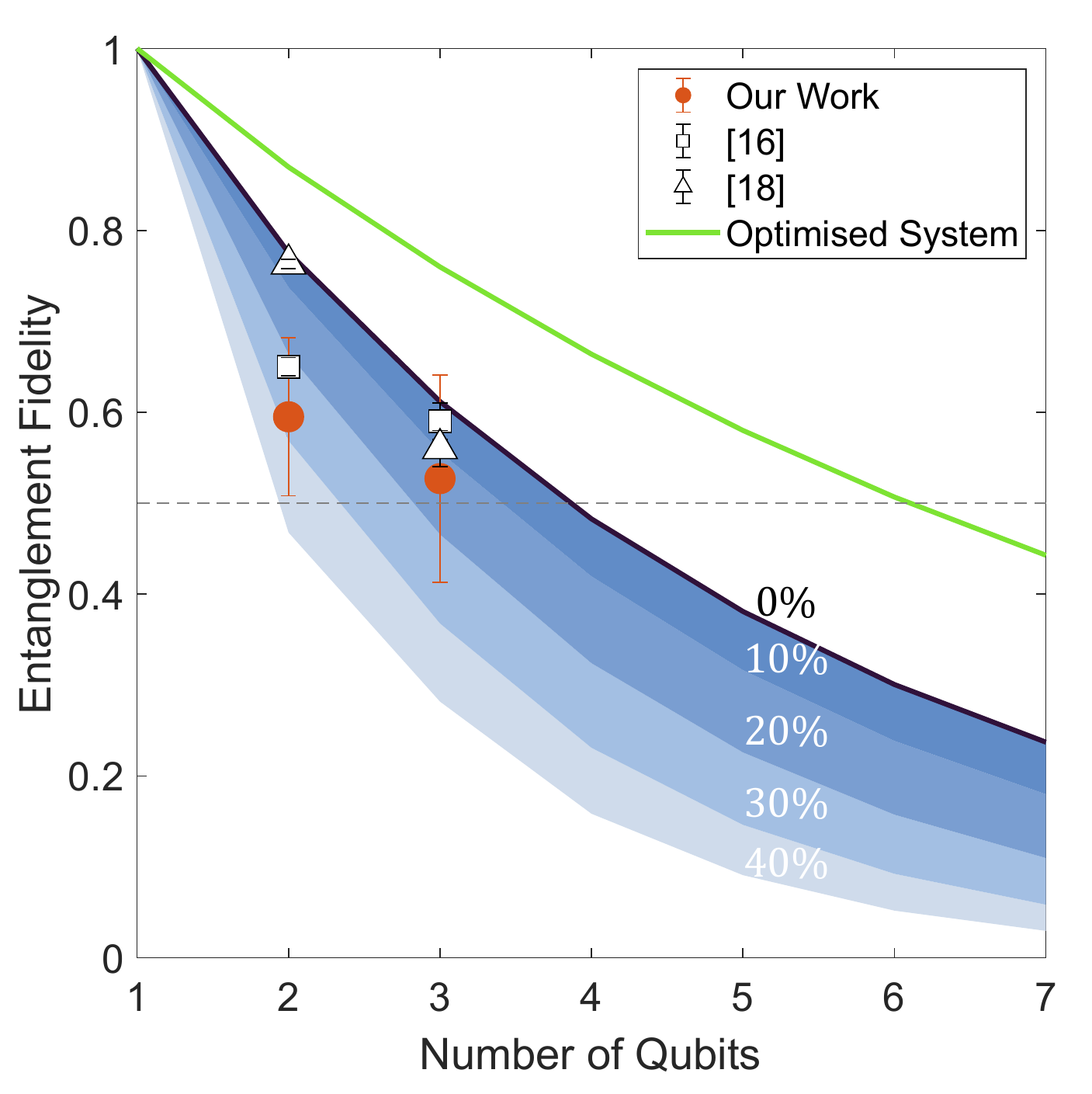} 
\caption{\textbf{Spin and $k$-photon entanglement fidelity} as a function of the total number of qubits (one stationary hole spin and $k$ emitted photons). The dashed horizontal line marks the 50 \% classical bound. Orange-filled circles show our experimental data; open squares and triangles are data from reference \cite{coste2023high} and \cite{meng2024deterministic}, respectively. The black solid line is the simulation of our current system, showing that a three-qubit state is achievable.  Beneath it, successive blue shaded bands trace out the simulated fidelity for a systematic error in excitation polarisation, magnetic field magnitude, and spin coherence, of $\epsilon=\{10 \%, 20 \%, 30 \%, 40 \%\}$ (darkest to lightest). The light green solid line shows the projected performance of our optimised QD system with Purcell-enhanced emission and extended spin coherence.}
\label{fig:Outlook} 
\end{figure}

For our simulations, we first adopt parameters drawn directly from the experiment; hole and electron spin coherence times of 4.8 ns and 0.8 ns, respectively, a trion lifetime of 0.8 ns and and a 300 ps post-selection window. We first assume perfect H-polarisation alignment of the excitation laser given by the QD coordinate system (see Supplementary Information Section 5). The resulting fidelities of the spin--k-photon state, with the hole spin as the first qubit, are $F=\{77.6 \%, 61.2 \%, 48.2 \% \}$ for $k={1,2,3}$, respectively, indicated by the black solid curve in \autoref{fig:Outlook}. The simulation also predicts a non-classical three-qubit state (2 photons, 1 spin) fidelity, for a system with no experimental errors. 

Comparing the simulation to our experimental data points (orange circles), it becomes clear that the current experiment does not reach its full potential. In practice, a host of imperfections degrade our multi‐photon fidelities: misalignment of both the excitation and detection polarisation (especially in the linear H/V basis, which also shifts the optimal pulse spacing $t_{12}$ and $t_{23}$), slight deviation in the magnitude of the magnetic field, a tilt of the magnetic field away from the QD’s crystallographic axis \cite{ramesh2025impact}, small errors in the inter‐pulse delays from splicing imperfections and the finite width of our 300-ps post-selection window (both of which disrupt the intended $\pi$/2 Larmor precession). 
Among these, we simulate the combined effect of excitation polarisation misalignment, magnetic field magnitude variation, and reduced spin coherence.

We find that as we deviate from the optimal H-polarised excitation, the pulse imprints an unwanted phase on the excited state qubit that cascades through each subsequent photon emission. Moreover, even mT-scale offsets in the applied magnetic field shift the spin’s precession period away from the exact $\pi$/2 rotation required for the implementation of a Hadamard gate. Similarly, reduced spin coherence (both for the electron and heavy hole), significantly degrades the overall fidelity. To illustrate these effects, we overlay a series of shaded blue bands beneath the ideal curve in \autoref{fig:Outlook} to trace out the achievable fidelity for increasing systematic error, $\epsilon=\{10 \%, 20 \%, 30 \%, 40 \%\}$. As it can be seen, the expected impact of this error on fidelity is indeed quite drastic - even a 20$\%$ error prevents three-qubit entanglement. Although we cannot attribute all of our current infidelity to these three sources of error alone, we avoid the complexity of considering contributions from the entire multi-dimensional parameter space of error sources, which may obscure rather than clarify the dominant effects. By focusing on a few, well-understood imperfections, the shaded bands thus serve as a practical way to visualise how even small systematic errors lead to noticeable drops in fidelity, underscoring the system’s sensitivity.

To contextualise our results, in \autoref{fig:Outlook} we compare the entanglement fidelities obtained in this work with those from previous experiments using InAs QDs emitting around 900 nm \cite{coste2023high, meng2024deterministic}. Some relevant results are excluded from the comparison due to incompatible entanglement metrics \cite{schwartz2016deterministic, cogan2023deterministic}. While our fidelities fall within the error bars of earlier works, they do not yet reach the highest values achieved at shorter wavelengths. One key factor limiting performance in our device is the significantly longer excited-state lifetime. Although we employ temporal post-selection to mitigate the excited spin precession during photon emission, this comes at the cost of a reduced entanglement generation rate and does not remove the constraint on the minimum pulse separation. Embedding QDs into optical resonators offers a promising solution by enhancing the spontaneous emission rate via the Purcell effect, thus eliminating the need for temporal post-selection. In particular, bullseye resonators have shown great potential, with simulations predicting Purcell factors up to 20 \cite{Barbiero.2022, Rickert.2023}, and experimental demonstrations achieving values of 3--5 in the telecom C-band \cite{Nawrath.2023, Joos.2024, Holewa.2024, Kim.2025, barbiero2025circular}.
Another important limitation is the relatively short spin coherence time, which could be extended by over an order of magnitude through control of the nuclear spin ensemble \cite{Gangloff.2019, zaporski2023ideal}. Additionally, due to the g-factor anisotropy, careful optimisation of the magnetic field angle with respect the QD's crystallographic axis, in combination with a coordinated optimisation of the excitation laser polarisation, has been shown to significantly enhance entanglement fidelity \cite{ramesh2025impact}.
Such enhancements are within practical reach.

Looking ahead, to assess the potential of our telecom QD interface after realistic improvements, we performed additional simulations assuming an excited state lifetime of 200 ps, corresponding to a Purcell factor of $F_p \sim 4$. We have also considered an improved hole spin coherence time of $T_2^* = 10$ ns, which is a conservative value compared to those reported for InAs QDs under nuclear spin narrowing \cite{meng2024deterministic}. As shown by the light green solid line in \autoref{fig:Outlook}, these moderate improvements, well within experimental reach, would enable the generation of entangled strings of up to six qubits, making telecom QDs competitive with their 900 nm counterparts. Together with the results demonstrated in this work, such advancements establish a clear path toward solid-state deterministic sources of entangled photons with direct emission in the telecom C-band, offering new opportunities for the implementation of practical fibre-based quantum networks and all-optical quantum repeaters.

\section*{Methods}
\textbf{Sample description:} 

The InAs/InP quantum dots are embedded within a distributed Bragg reflector cavity with a cubic zirconia solid immersion lens on top to enhance the photon collection efficiency into the microscope. Further description in Supplementary Information Section 1.

\noindent \textbf{Experimental setup: } 

To enable phonon-assisted excitation, we shape the output of a broadband pulsed laser. Following initial spectral filtering, the laser pulses are temporally stretched to 15.8 ps using a 4-f spectral filter. The stretched pulses are then passed through a series of fibre beamsplitters to create the ideal excitation sequence (see Supplementary Information Section 2). A blue detuning of 1.5 nm from the X$^{+}$ transition is employed as an optimal trade-off between minimizing laser leakage through the spectral filters and maintaining sufficient power for excitation of the QD. To isolate the QD emission, a cascade of three high-extinction band-pass optical filters are included in the collection path of the microscope.

For synchronisation, the fixed 80 MHz repetition rate of our excitation laser is used as a reference clock for the rest of the experimental apparatus. First, this reference signal seeds a phase locked loop, from which a 40 MHz sync output is obtained. This 40 MHz signal is distributed to an arbitrary wavefunction generator, the above-band laser, and the timetagger.

\section*{Data Availability }
The authors declare that the data supporting the findings of this study are available from the corresponding authors upon reasonable request.

\bibliographystyle{naturemag} 
\bibliography{spinphoton}

\section*{Acknowledgements}
The authors gratefully acknowledge the usage of wafer material developed during earlier projects in partnership with the National Epitaxy Facility at the University of Sheffield. They further acknowledge funding from the Ministry of Internal Affairs and Communications, Japan, via the project 'Research and Development for Construction of a Global Quantum Cryptography Network' in 'R\&D of ICT Priority Technology' (JPMI00316).  P. L. gratefully acknowledges funding from the Engineering and Physical Sciences Research Council (EPSRC) via the Centre for Doctoral Training in Connected Electronic and Photonic Systems, grant EP/S022139/1. \\

\section*{Author Contributions Statement}
D.A.R, A.J.S, T.M, and R.M.S supervised the project. T.M and R.M.S guided the experiment. G.S fabricated the emission enhanced QD structure. P.L, J.H, and G.S took the measurements and analysed the data. P.L and M.W worked on the theoretical modelling of the master equation and process tomography. A.B worked on the phenomenological model. P.L, J.H, G.S, A.B, and T.M wrote the manuscript. All authors discussed the results and commented on the manuscript. \\

\section*{Competing Interests Statement}
The authors declare that they have no competing financial interests.

\end{document}